\documentclass[usenatbib]{mnras}
\usepackage{graphicx}
\usepackage{graphics}
\usepackage{amsmath}
\usepackage{multirow}
\usepackage{color}
\usepackage{xcolor}
\usepackage{bm}		
\usepackage{float}
\usepackage{rotating}
\usepackage{adjustbox}
 \usepackage{gensymb}

\title[]{The Tale of the Milky Way Globular Cluster NGC 6362 - I. The Orbit and its possible extended star debris features as revealed by {\it Gaia} DR2}

\author[Kundu et al.]{
	Richa Kundu$^{1}$\thanks{E-mail: richakundu92@gmail.com},
	Jos\'{e} G. Fern\'{a}ndez-Trincado,$^{2,3}$\thanks{E-mail: jfernandezt87@gmail.com or jfernandez@obs-besancon.fr},
	Dante Minniti$^{4,5,6}$
	\newauthor
	Harinder P. Singh$^{1}$,
	Edmundo Moreno$^{7}$,
	C\'eline Reyl\'e$^{3}$,
    Annie C. Robin$^{3}$,
	\newauthor
	and
	Mario Soto$^{2}$
	\\
	$^{1}$Department of Physics and Astrophysics, University of Delhi, Delhi-110007, India\\
	$^{2}$Instituto de Astronom\'ia y Ciencias Planetarias, Universidad de Atacama, Copayapu 485, Copiap\'o, Chile\\
	$^{3}$Institut Utinam, CNRS UMR 6213, Universit\'e Bourgogne-Franche-Comt\'e, OSU THETA Franche-Comt\'e, Observatoire de Besan\c{c}on, \\ BP 1615, 25010 Besan\c{c}on Cedex, France\\
	$^{4}$Instituto Milenio de Astrofisica, Santiago, Chile\\
	$^{5}$Departamento de Ciencias Fisicas, Facultad de Ciencias Exactas, Universidad Andres Bello, Av. Fernandez Concha 700, Las Condes, \\Santiago, Chile\\
	$^{6}$Vatican Observatory, V00120 Vatican City State, Italy\\
	$^{7}$Instituto de Astronom\'ia, Universidad Nacional Aut\'onoma de M\'exico, Apdo. Postal 70264, M\'exico D.F., 04510, M\'exico\\
}
 
\begin{document}
\label{firstpage}
\pagerange{\pageref{firstpage}--\pageref{lastpage}}
\maketitle

\begin{abstract}	
{We report the identification of possible extended star debris candidates beyond the cluster tidal radius of NGC 6362 based on the second {\it Gaia} data release ({\it Gaia} DR2). We found 259 objects possibly associated with the cluster lying in the vicinity of the giant branch and 1--2 magnitudes fainter/brighter than the main-sequence turn-off in the cluster color-magnitude diagram and which cover an area on the sky of $\sim$4.1 deg$^{2}$ centered on the cluster. We traced back the orbit of NGC 6362 in a realistic Milky-Way potential, using the \texttt{GravPot16} package, for 3 Gyrs. The orbit shows that the cluster shares similar orbital properties as the inner disk, having peri-/apo-galactic distances, and maximum vertical excursion from the Galactic plane inside the corotation radius (CR), moving inwards from CR radius to visit the inner regions of the Milky Way. The dynamical history of the cluster reveals that it has crossed the Galactic disk several times in its lifetime and has recently undergone a gravitational shock, $\sim 15.9$ Myr ago, suggesting that less than 0.1\% of its mass has been lost during the current disk-shocking event. Based on the cluster's orbit and position in the Galaxy, we conclude that the possible extended star debris candidates are a combined effect of the shocks from the Galactic disk and evaporation from the cluster. Lastly, the evolution of the vertical component of the angular momentum shows that the cluster is strongly affected dynamically by the Galactic bar potential.
}
\end{abstract}                
 
\begin{keywords}
Galaxy: kinematics and dynamics – (Galaxy): globular clusters: individual: NGC 6362
\end{keywords}
 

\section{Introduction}

Extra-tidal stellar material associated with globular clusters  is spectacular evidence for satellite disruption at the present day, which provides significant clues about the dynamical history of the clusters and their host galaxies. Globular clusters evolve dynamically under the influence of the gravitational potential well of their host galaxy \citep{gnedin97, murali1997, Leon2000, Kunder18, Minniti18}, resulting in the escape of the stars close to the tidal boundary of the cluster, consequently forcing the cluster cores to contract and envelopes to expand \citep[e.g.,][]{Leon2000, kunder2014}. Therefore, globular clusters are important stellar systems to study the evolution, structure and dynamics of their host galaxy. 

Globular clusters lose stars mainly due to dynamical processes like dynamical friction, tidal disruption, bulge and disk shocking and evaporation \citep{fall77, fall85}. Dynamical friction is due to the gravitational pull of the field stars that are accumulated behind the cluster motion. These stars slow down the cluster and pull some of the loosely bound stars away from it. This effect is more pronounced in the bulge of the Galaxy where the density of field stars is higher. Dynamical friction has been proposed in many studies \citep{chandrasekhar43, mulder83, white83, tremaine84, capdol05b, TR, arca14} but the observational evidence has been more elusive, while tidal disruption have been observed \citep{Leon2000, Odenkirchen2001, Belokurov2006, Grillmair2006, Grillmair10, ostholt10, jordi2010, sollima11, balbinot11, kuzma15, myeong17, navarrete17} and studied by many \citep{king62, tremaine75, chernoff86, capdol93, weinberg94, meylan97, gnedin97, vesperini97, combes99, lotz01, capdel05, majewski12, kupper12, Majewski12a, flores12, knierman13, mulia14, hozumi15, rodruck16, Fernandez-Trincado2013, Fernandez-Trincado2015a, Fernandez-Trincado2015b, Fernandez-Trincado2016a, Fernandez-Trincado2016b, Fernandez-Trincado2017a, Fernandez-Trincado2017b, Fernandez-Trincado2017c, balbinot18, Myeong18, Kundu2019, Mackereth2019}. 

NGC 6362 is a nearby low mass globular cluster with intermediate metallicity, located in the bulge/disk of the Milky Way galaxy \citep{carretta10}. It has an age of $\sim$12.5$\pm$0.5 Gyr, which is enough to evolve under the gravitational potential of the Milky Way. Therefore, identifying possible tidal tails around NGC 6362 is especially intriguing to study the cluster dynamics in the bulge/disk region, which is poorly understood. Recently, \citet{baumgardt18} presented a catalog of masses, structural profiles and velocity dispersion values for many Galactic globular clusters including NGC 6362. They found that this cluster fits a King profile with a constant velocity dispersion as  a function of radius, hence there was no evidence of a tidal tail. However, their measurements were concentrated to the inner regions, extending only out to 400 arc-sec away from the center. 

In the present work, we report the detection of potential extended star debris associated with NGC 6362. We have taken advantage of the exquisite data from {\it Gaia} Data Release 2 \citep[Gaia DR2,][]{gaiadr2} to search for such extended star debris features around NGC 6362. To give a proper explanation for the presence of the observed possible star debris, we time-integrated backward the orbit of NGC 6362 to 3 Gyrs under variations of the initial conditions (proper motions, radial velocity, heliocentric distance, Solar position, Solar motion and the velocity of the local standard of rest) according to their estimated errors. Our analysis indicates that the cluster is dynamically affected by the Galactic bar potential, presently experiencing a bulge/bar shocking, with considerable amount of mass loss, which can be observed as stars present in the immediate neighborhood of the cluster. A similar analysis was recently carried out by \citet{Minniti18} for NGC 6266 (also known as M62) using extra-tidal RR Lyrae stars.

This paper is organized as follows. In Section \ref{section2}, we select the possible star debris candidates beyond the cluster tidal radius of NGC 6362. In Section \ref{contamination},  we discussed the significance of the observed star debris. In Section \ref{section3}, we determine its most likely orbit using novel galaxy modeling software called \texttt{GravPot16}. In Section \ref{mass_loss}, we discussed the mass lost by the cluster due to various processes. The concluding remarks are summarised in Section \ref{section4}.

\begin{table}
			\begin{center}
	\setlength{\tabcolsep}{1.5mm}  
	\caption{NGC 6362 --Sun parameters.}
	\begin{tabular}{lcc}
			\hline
			\hline
			Parameter & Value & Reference \\
			\hline
			\hline
			NGC 6362 &  & \\
			\hline
			\hline
			$\alpha$ ($^{\circ}$), $\delta$ ($^{\circ}$) &  262.979, $-$67.048  &   (a) \\
			Distance (kpc) & 7.6          &    (a) \\
			$R_{gal}$ (kpc) & 4.71   &  \\
 			$\mu_{\alpha}$ (mas/yr) & -5.507$\pm$0.052   &  (a) \\
			$\mu_{\delta}$ (mas/yr) & -4.747$\pm$0.052  &  (a) \\
			$V_{los}$ &   $-$14.58$\pm$0.18 &  (a) \\			
			Tidal Radius (pc) & 30.73     &  (b)  \\
			Mass (M$_{\odot}$)  & $\sim$ 10$^5$  &    (b) \\
			Metallicity & $-1.07$          &    (d)\\
			Age (Gyr)      & 12.5$\pm$0.5     &    (e)\\
			\hline
			\hline
			Sun &  &\\
			\hline
			\hline
		$R_{\odot}$ (kpc) & 8.3   &  (f) \\
		     $U_{\odot}, V_{\odot}, W_{\odot} $  (km s$^{-1}$)  &  11.10, 12.24, 7.25 & (f) \\
		     $V_{LSR}$   (km s$^{-1}$) & 239  & (f)  \\
			\hline
			\hline
	\end{tabular}  \label{tab:t1}\\
			\end{center}
				\raggedright {(a)  \citet{Vasiliev18}; (b) \citet{TR} ; (c) \citet{dalessandro14}; (d) \citet{massari17} ; (e) \citet{dotter2010}; (f) \citet{Brunthaler2011}. }
\end{table}

\begin{figure*}
	\begin{center}
		\includegraphics[width=1.0\textwidth,keepaspectratio]{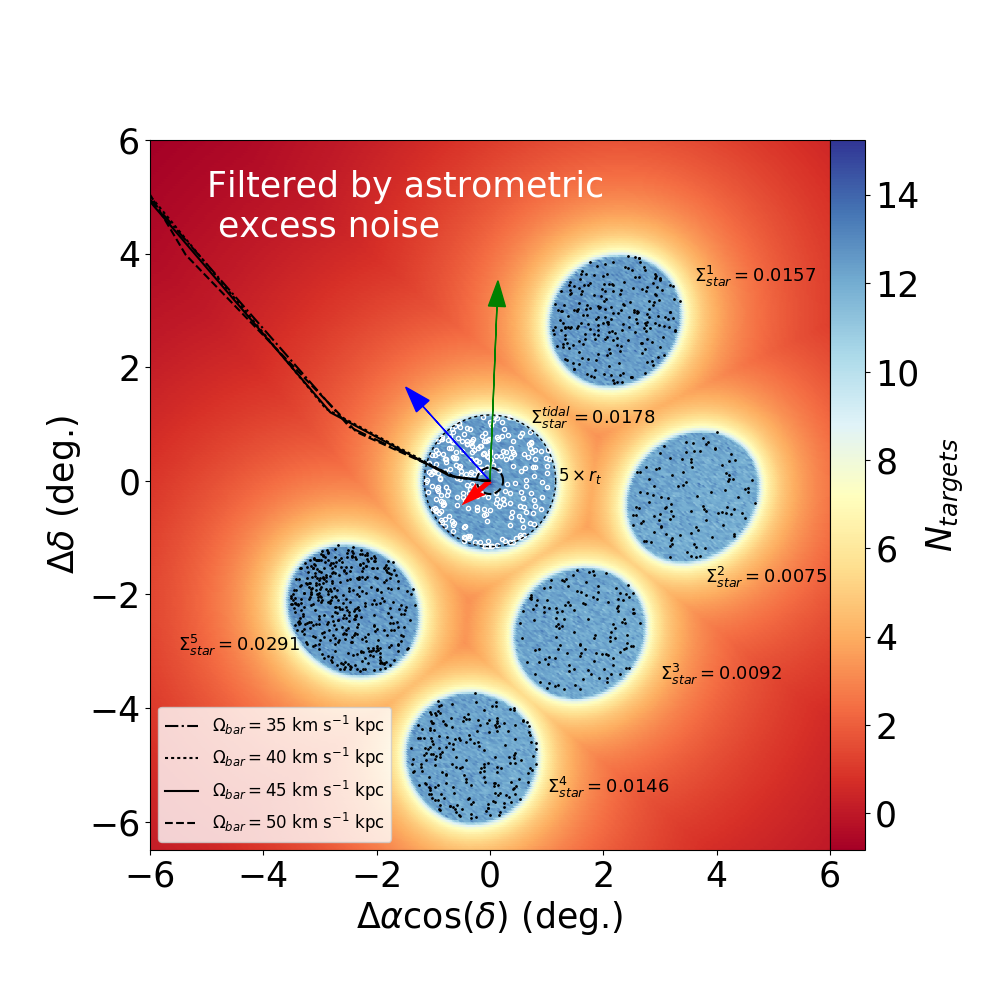}
		\caption{ The {\it Gaia} DR2 positions for the highest likelihood star debris candidates in the region of NGC 6362 shown with unfilled white circles. The inner and outer black dashed circles are the tidal radius ($r_t$) and 5$\times r_t$, respectively (see text). The arrows indicate the directions of the cluster proper motion (red arrow), with a preferential direction toward S--W, the Galactic center (G.C.--green arrow), and the direction perpendicular to the galactic plane (blue arrow). The computed orbit (black lines) of the cluster is displayed assuming four different values of the bar patterns speed (35, 40, 45, and 50 km s$^{-1}$ kpc) in the \textit{GravPot16} package (see text). Five adjacent regions containing field stars (foreground and background) whose proper motions and distribution in the CDM are overlapped with cluster members, and in which the contamination was evaluated. The expected surface density of potential members and each adjacent field is internally indicated, which overlap all the criteria adopted in this work.}
		\label{fig:f3}
	\end{center}
\end{figure*}

\section{Identification of extended star debris candidates around NGC 6362}
\label{section2}

To search for the extended star debris features around the cluster NGC 6362, we have made use of the second {\it Gaia} data release \citep[Gaia DR2,][]{gaiadr2}. We first download {\it Gaia} DR2 in a cone around the cluster with radius around five tidal radii where we tried to identify the star debris, which contains 276,391 objects.

Since NGC 6362 is relatively far, we decided to pay particular attention to avoid contamination by data processing artifacts and/or spurious measurements. Therefore, we adopted the following conservative cuts on the columns of the {\it Gaia} DR2 \texttt{GAIA\_SOURCE} catalogue:

\begin{itemize}
\item[(i)]   \texttt{ASTROMETRIC\_GOF\_AL $<$ 3}. This cut ensures that the statistics astrometric model resulted in a good fit to the data;
\item[(ii)]  \texttt{ASTROMETRIC\_EXCESS\_NOISE\_SIG $\leq$ 2.} This criterion ensured that the selected stars were astrometrically well-behaved sources;
\item[(iii)] \texttt{$-$0.23 $\leq$ MEAN\_VARPI\_FACTOR\_AL $\leq$ 0.32 AND} \texttt{VISIBILITY\_PERIODS\_USED > 8}. These cuts were used to exclude stars with parallaxes more vulnerable to errors;
\item[(iv)] \texttt{G $<$ 19 mag}. This criterion minimized the chance of foreground contamination.
\end{itemize}

 Here we only give a rough overview and refer the reader to \citet{Marchetti18} for a detailed description of these high-quality cuts.

 The final sample so selected amounts to a total of 83,406 stars. From this sample, we further retain as candidate members of the cluster those objects which lie in an annular region around the cluster with its inner radius as the tidal radius \citep[$r_t=$13.907 arcmin;][]{TR} of NGC 6362, and an outer radius equal to 5 times its tidal radius, as displayed in Figure \ref{fig:f3}. This reduces our sample to 77,549 objects.

As a consistency check, to verify the validity of highest likelihood star debris candidates based on their position on the sky only, the sample was restricted to the stars whose proper motions match with the proper motion of the cluster within 3$\sigma_{\mu}$, where $\sigma_{\mu}$ is the total uncertainty in quadrature obtained from a 2-dimensional Gaussian fit. For this purpose, a 2-dimensional Gaussian smoothing routine was applied in proper motion space for stars with $G<19$ mag within 2$\times$r$_{half-mass}$ from the centre of the cluster. A 2D Gaussian was fitted to this sample and membership probabilities are assigned. With this procedure, we found $\mu^{2D}_{\alpha} \pm \sigma_{\alpha} = -5.511 \pm 0.237$ mas yr$^{-1}$ and $\mu^{2D}_{\delta} \pm \sigma_{\delta} = -4.742 \pm 0.302$ mas yr$^{-1}$, and $\sigma_{\mu} = 0.38$ mas yr$^{-1}$, our results also agree remarkably well with the more recent measurements of PMs for NGC 6362, e.g.: $\mu_{\alpha} = -5.507 \pm 0.052$ mas yr$^{-1}$, and $\mu_{\delta} = -4.747 \pm 0.052$ from \citet{Vasiliev18}. A star was considered to be a GC member if its proper motion differs from that of NGC 6362 by not more than 3$\sigma_{\mu}$, leaving us with a grand total of 1,503 stars. The content of nearby stars in our initial sample is reduced by excluding those objects with estimated distances from \citet{jones18} confined to a sphere of radius 3 kpc around the Sun. This cut is motivated by the fact that at large latitudes away from the disk, the priors would be expecting distant stars to be much closer to us than they truly are and force the stars towards these closer, unrealistic distances. Therefore, the distances from the Bailer-Jones's catalog should just be following the priors and would not account for distant over-densities. This reduces our sample to 826 objects.

\begin{figure*}
	\begin{center}
		\includegraphics[width=1.0\textwidth,keepaspectratio]{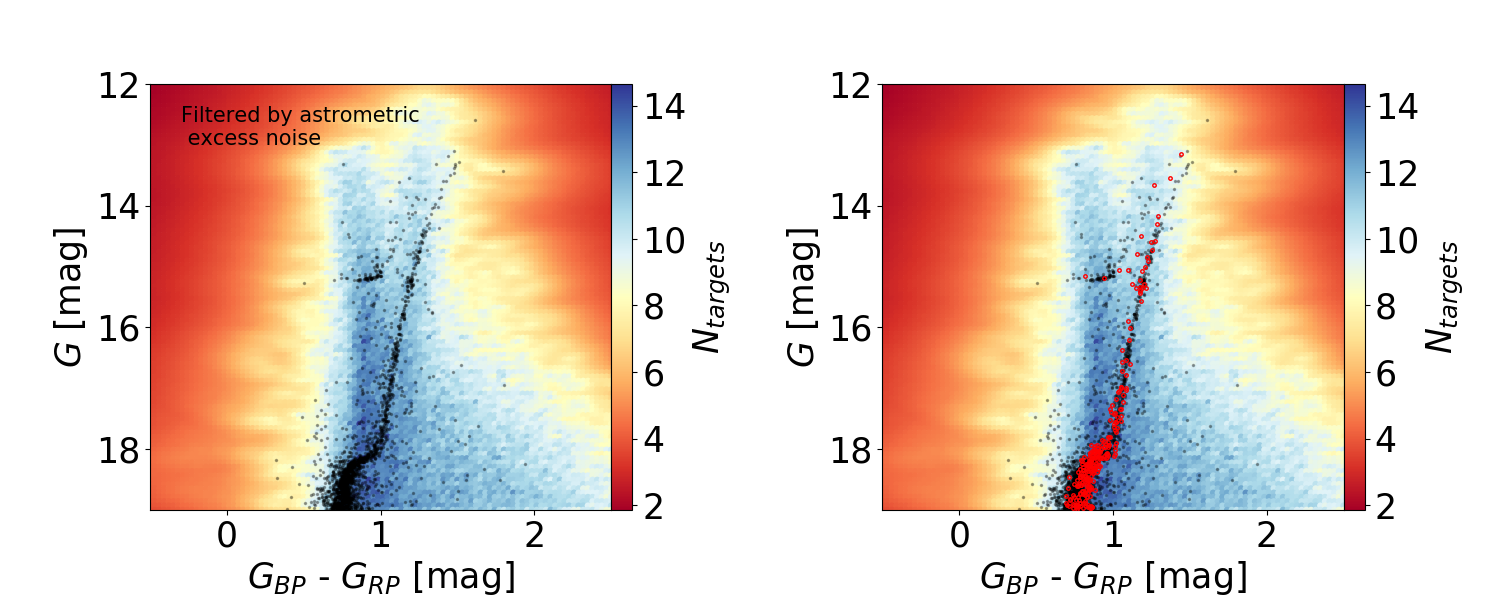}
		\includegraphics[width=1.0\textwidth,keepaspectratio]{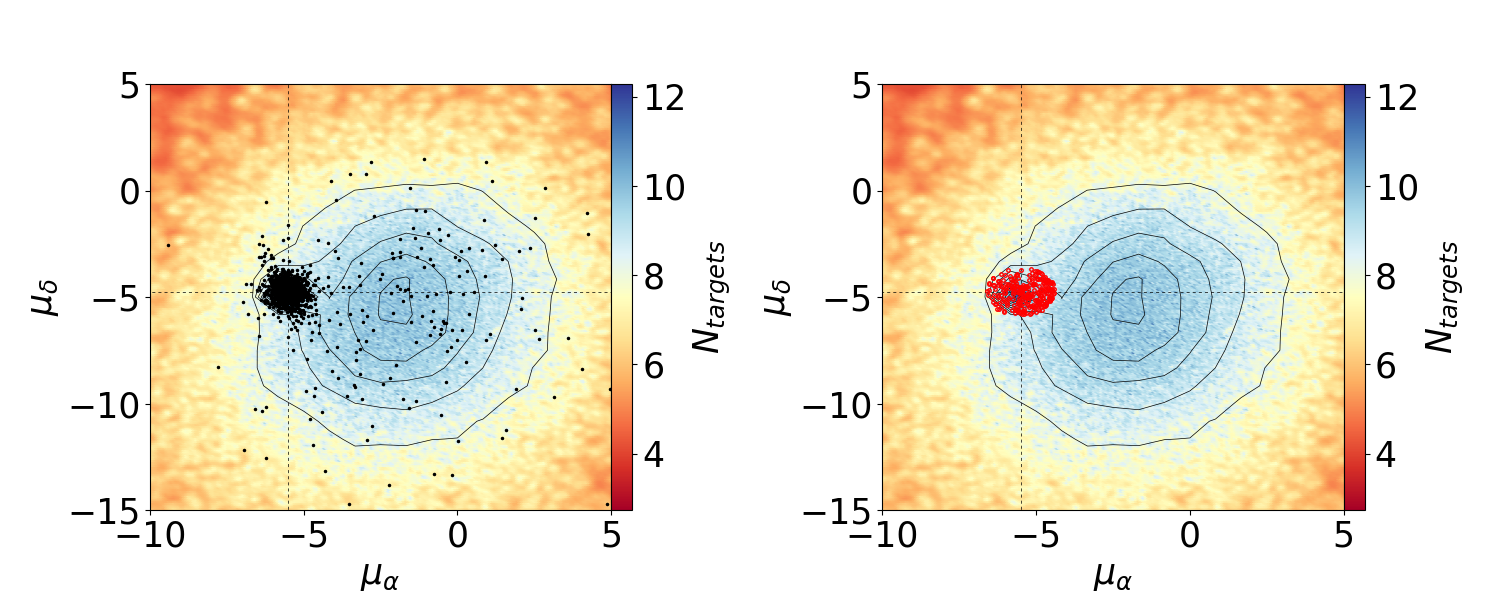}
		\caption{Kernel Density Estimate (KDE) smoothed distribution of the Colour-magnitude diagram of stars within $5\times{}r_{t}$ from the photometric centre of NGC 6362 (\textit{top rows}) and proper motions in the region of the cluster (bottom rows). Left panels illustrates the stars which pass the astrometric excess noise cutoffs, for stars in the field and stars within $2\times{}r_{half-mass} \sim 4.1$ arcmin from the centre of the cluster (black dots). Right panels illustrates the position in the CMD and VPD for the highest likelihood of possible extended star debris candidates (red dots). The black dashed lines show the nominal proper motion values for NGC 6362 at $\mu_{\alpha} = -$5.507 mas yr$^{-1}$, $\mu_{\delta} = -$4.747 mas yr$^{-1}$ \citep{Vasiliev18} while the white contour line encloses the density of fore/background stars and cluster itself.}
		\label{fig:f5}
	\end{center}
\end{figure*}

Thus we found a total of 826 possible star debris candidates of NGC 6362, which share an apparent proper motion close to the nominal value of the cluster, suggesting that these stars could possibly be evaporated material from NGC 6362. Therefore, to be sure that our candidate members are actually part of the cluster system, we selected those stars whose locations on the Colour-magnitude diagram (CMD) clearly lie on or near the prominent main branches of NGC 6362, as illustrated by the red symbols in figure~\ref{fig:f5}. A total of 259 possible extended star debris candidates passed these quality cuts as illustrated in Figure \ref{fig:f3} and Figure \ref{fig:f5} (highlighted by red symbols).

 To summarize, the possible star debris members of the cluster in Figure \ref{fig:f5} show the following: proper motions that are very concentrated as expected in the vector point diagram (hereafter VPD) of a globular cluster, and a CMD with the characteristic features of a globular cluster, e.g., the main sequence, the turn-off, the red giant branch and some stars in the horizontal branch. It is important to note that the determination of the possible extended star debris of NGC 6362 could include some field stars as members or vice versa, in \S\ref{contamination} we perform an estimation of the degree of contamination of the extracted members, i.e., the possible number of field stars that could have been labelled as possible extended star debris members of the cluster.

This finding gives possible clues about the recent dynamical history of  NGC 6362, which suggests that this cluster could eventually form tidal tails or could also be associated with the recent encounter of the cluster with the disk.

 Table~\ref{tab:tdata} lists the main parameters of the 259 possible extended star debris. Figure \ref{fig:f5} shows consistently the validity of our probable extended star debris members which share an apparent proper motion close to the value for NGC 6362, suggesting these stars are probable members of the cluster. 
 
It is important to note that most of the stars inside $2\times{}r_{half-mass}$ of the cluster are spread in spread in proper motions, as illustrated by black dots in Figure \ref{fig:f5}, consequently one may be lead to conclude that it is related to contamination by foreground/background stars which would seem to be the most likely explanation for the significantly higher proper motion values. Thus, we also expect that our sample may be significantly contaminated from other Galactic stellar populations (see \S\ref{contamination}). To alleviate this situation, a detailed chemical abundance analysis will be necessary to understand their relation, if any, with the cluster.

{
\section{Significance of the detection of possible extended star debris around NGC 6362}
\label{contamination}

It is important to note that the main tracers of the possible extended star debris of NGC 6362 identified in this work are main-sequence (MS) stars and subgiant stars 1--2 magnitudes fainter and brighter than the MS turn-off (TO), respectively. However, the cluster stars beyond cluster tidal radius are hidden in the CMD due to the combination of the contributions of a minor fraction of cluster members and fore-/back-ground stellar populations from the different Milky Way components (mainly the thin-/thick disk, and halo).

In this sense, we attempt to estimate the significance of the detection in our photometry and PMs space. For this purpose, we have compared the observed stellar counts with those computed from the synthetic CMDs generated with the updated version of the Besan\c{c}on Galaxy model for the same line-of-sight and solid angle, after correcting for completness. For a more detailed description of the Besan\c{c}on Galaxy model, we refer the readers to \citet[][the full, basic description]{Robin2003}, \citet[][update on the thick disc]{Robin2014}, \citet[][update on kinematics]{Robin2017} and \citet[][update on the stellar evolutionary models]{Lagarde2017}. The observed stars considered to derive the significance of a subjacent population are those contained in the CMD and PM space as illustrated in Figure \ref{fig:f5}.

We calculated the expected number of Milky Way stars over the survey area and in distance range D$_{\odot}$ $>$3 kpc from the Besan\c{c}on Galaxy model. We found $N_{\rm model} \sim 167 \pm 13$ stars in the area of the {\it Gaia} footprint around NGC 6362. The cited error is Poisson statistics. We can then estimate the significance of the detection with respect to the synthetic model in the following manner: $\delta \approx (N_{\rm model} - N_{\rm extra-tidal}) / (N_{\rm model} + N_{\rm extra-tidal})^{1/2}$, where $N_{\rm extra-tidal}$ is the number of observed stars following the criteria described above. We obtain a $\delta \sim 4.5$ detection above the foreground and background population. 

{ 
Another way to perform an estimation of the degree of contamination of the extracted members relies in upon apply our method in adjacent regions (defined with the same area than our explored region) around the cluster, as illustrated in Figure \ref{fig:f3}. Performing an analysis like that mentioned in the beginning of \S\ref{section2} but counting all the stars in the field instead of only those potential members around the cluster, we obtain rough estimates of the expected contamination in our sample. We note that the incompleteness of the {\it Gaia} DR2 catalogue itself has not been taken into account in our computations, therefore, our estimates are upper limits to the actual completeness for the most favorable cases (low-density fields). Figure \ref{fig:f3} the expected surface density ($\Sigma^{1}_{star}$, $\Sigma^{2}_{star}$, $\Sigma^{3}_{star}$, $\Sigma^{4}_{star}$, and $\Sigma^{5}_{star}$) of foreground/background stars (black dots) in five adjacent regions around NGC 6362. Those densities remain low as compared to our potential sample, with the exception of $\Sigma^{5}_{star} = 0.0291$, which is higher due to that this region lies in the direction of the sky containing the highest densities of field stars, for this reason we have also avoid additional adjacent regions toward the direction North-West of the cluster. Finally, based on $\Sigma^{1}_{star}$, $\Sigma^{2}_{star}$, $\Sigma^{3}_{star}$ and $\Sigma^{4}_{star}$ we estimate the degree of contamination, i.e., the fraction of field stars that could have been erroneously labelled as possible extended star debris members, which is expected that $\sim$40\% ($\sim$ 103$\pm$10 stars) to 80\% ($\sim$207$\pm$14 stars) of the field stars could have been erroneously extracted as members in our sample (which we call contamination of the members). This rough estimation, point-out a good agreement between the Besan\c{c}on Galaxy model and the data in the degree of contamination of the extracted members by other Galactic stellar populations. In both cases, a future inventory of the chemistry of these stars, in particular the elements involved in the proton-capture reactions (i.e, C, N, O, Mg, Al, among other) will be crucial to confirm or refute the cluster nature of these star debris candidates in a similar fashion as \citet{Fernandez-Trincado2016a, Fernandez-Trincado2017c, Fernandez-Trincado2018, Fernandez-Trincado2019a, Fernandez-Trincado2019b, Fernandez-Trincado2019d}. These stars will be later analyzed using high-resolution ($R\sim22,000$) spectra from the APOGEE-2S survey \citep{Majewski2016, Zasowski2017} in order to investigate its chemical composition. 
}

\section{The Orbit of NGC 6362}
\label{section3}

We estimated the probable Galactic orbit for NGC 6362 in order to provide a possible explanation to the possible extended star debris identified in this work. For this, we used a state-of-the art orbital integration model in an (as far as possible) realistic gravitational potential, that fits the structural and dynamical parameters of the galaxy to the best we know of the recent knowledge of the Milky Way. For the computations in this work, we have employed the rotating "boxy/peanut" bar model of the novel galactic potential model called \texttt{GravPot16}\footnote{\url{https://gravpot.utinam.cnrs.fr}} along with other composite stellar components. The considered structural parameters of our bar model, e.g., mass, present-day orientation and pattern speeds, are within observational estimations: 1.1$\times$10$^{10}$ M$_{\odot}$, 20$^{\circ}$ and 35 to 50 km s$^{-1}$ kpc, respectively. The density-profile of the adopted "boxy/peanut" bar is exactly the Model-S as in \citet{Robin2012}, while the mathematical formalism to derive a correct global gravitational potential of this component will be explained in a forthcoming paper (Fern\'andez-Trincado et al. 2019, in preparation).

\texttt{GravPot16} considers on a global scale a 3D steady-state gravitational potential for the Galaxy, modelled as the superposition of axisymmetric and non-axysimmetric components. The axisymmetric potential is made-up of the superposition of many composite stellar populations belonging to seven thin disks following the Einasto density-profile law \citep{Einasto1979}, superposed along with two thick disk components, each one following a simple hyperbolic secant squared decreasing vertically from the Galactic plane plus an exponential profile decreasing with Galactocentric radius as described in \citet{Robin2014}. We also implemented the density-profile of the interstellar matter (ISM) component with a density mass as presented in \citet{Robin2003}. The model, also correctly accounts for the underlying stellar halo, modelled by a Hernquist profile as already described in \citet{Robin2014}, and surrounded by a single spherical Dark Matter halo component \citet{Robin2003}. Our dynamical model has been adopted in a score of papers \citep[e.g.,][]{Fernandez-Trincado2016a, Fernandez-Trincado2016b, Fernandez-Trincado2017a, Fernandez-Trincado2017b, Fernandez-Trincado2017c, Fernandez-Trincado2018, Fernandez-Trincado2019a, Fernandez-Trincado2019b, Fernandez-Trincado2019c, Fernandez-Trincado2019d, Robin2017}. For a more detailed discussion, we refer the readers to a forthcoming paper (Fern\'andez-Trincado et al., in preparation).

For reference, the Galactic convention adopted by this work is: $X-$axis is oriented toward $l=$ 0$^{\circ}$ and $b=$ 0$^{\circ}$, and the $Y-$axis is oriented toward $l$ = 90$^{\circ}$ and $b=$ 0$^{\circ}$, and the disk rotates toward $l=$ 90$^{\circ}$; the velocity components are also oriented along these directions. In this convention, the Sun's orbital velocity vector is [U$_{\odot}$,V$_{\odot}$,W$_{\odot}$] = [$11.1$, $12.24$, $7.25$] km s$^{-1}$ \citep{Brunthaler2011}. The model has been rescaled to the Sun's galactocentric distance, 8.3 kpc, and the local rotation velocity of $239$ km s$^{-1}$. 

 For the computation of the Galactic orbits for NGC 6362, we have employed a simple Monte Carlo scheme for the input data listed in Table \ref{tab:t1}, and the Runge-Kutta algorithm of seventh-eight order elaborated by \citet{fehlberg68}. The uncertainties in the input data (e.g., distance, proper motions and line-of-sight velocity errors), were propagated as 1$\sigma$ variations in a Gaussian Monte Carlo re-sampling in order to estimate the more probable regions of the space, which are crossed more frequently by the simulated orbits as illustrated in Figure \ref{fig:f5}. The error bar for the heliocentric distance is assumed to be 1 kpc. We have sampled half million orbits, computed backward in time during 3 Gyr. Errors in the calculated orbital elements were estimated by taking half million samples of the error distributions and finding the 16$^{th}$ and 84$^{th}$ percentiles as listed in Table \ref{tab:t3}. The average value of the orbital elements was found for half million realizations, with uncertainty ranges given by the 16th and 84th percentile values, as listed in Table \ref{tab:t3}, where r$_{\rm peri}$ is the average perigalactic distance, r$_{\rm apo}$ is the average apogalactic distance and Z$_{\rm max}$ is the average maximum distance from the Galactic plane.

	Figure \ref{fig:f9} shows the probability densities of the resulting orbits projected on the equatorial (left column) and meridional (right column) Galactic planes in the non-inertial reference frame where the bar is at rest. The orbital path (adopting central values) is shown by the black line in the same figure. The green and yellow colors correspond to more probable regions of the space, which are crossed more frequently by the simulated orbits. We found that most of the simulated orbits are situated in the inner bulge region, which means that NGC 6362 is on high eccentric orbit (with eccentricities greater than 0.45) reaching out to a maximum distance from the Galactic plane larger than 2 kpc with a perigalacticon of $\sim$ 2 kpc and an apogalactic distance of $\sim 6$ kpc. On the other hand, NGC 6362 orbits have energies allowing the cluster to move inwards from the bar's corotation radius ($<$ 6.5 kpc). In this region a class of orbits appears around the Lagrange points on the minor axis of the bar that can be stable and have a banana-like shape parallel to the bar (see lower panel with $\Omega_{bar}=50$ km s$^{-1}$ kpc in Figure \ref{fig:f9}), while the Lagrange orbits libating around Lagrange points aligned with the bar are unstable and are probably chaotic orbits. Our model naturally predicts trajectories indicating that NGC 6362 is confined to the inner-disk.

Additionally, in figure~\ref{fig:f8} we show the variation of the z-component of the angular momentum in the inertial frame, $L_{z}$, as a function of time and $\Omega_{\rm bar}$. Since, this quantity is not conserved in a model like \texttt{GravPot16} (with non-axisymmetric structures), we follow the change, \{-$L_{z}$,+$L_{z}$\}, where negative $L_z$ in our reference system means that the cluster orbit is prograde (in the same sense as the disk rotation). Both prograde and prograde-retrograde orbits with respect to the direction of the Galactic rotation are clearly revealed for NGC 6362. This effect is strongly produced by the presence of the galactic bar, further indicating a chaotic behavior.

\begin{table}
	\centering
	\caption{Orbital parameters of NGC 6362, with uncertainty ranges given by the 16$^{th}$ (subscript) and 84$^{th}$ (superscript) percentile values.
	}
	\label{tab:t3}
	\begin{tabular}{ccccc} 
		\hline
		$\Omega_{\rm bar}$     &   r$_{\rm peri}$       &    r$_{\rm apo}$       &  Z$_{\rm max}$         &    eccentricity      \\ 
		km s$^{-1}$ kpc$^{-1}$   &      kpc               &       kpc              &     kpc                &                      \\
		\hline
		
		35            &  2.02$^{2.18}_{1.81}$  &   5.29$^{5.68}_{5.03}$ &  3.41$^{3.83}_{3.30}$  &  0.45$^{0.49}_{0.42}$ \\ 
		40            &  1.98$^{2.17}_{1.87}$  &   5.38$^{6.03}_{5.16}$ &  3.45$^{3.84}_{3.19}$  &  0.47$^{0.49}_{0.44}$ \\ 
		45            &  2.04$^{2.22}_{1.97}$  &   5.94$^{6.89}_{5.72}$ &  3.55$^{4.14}_{3.14}$  &  0.49$^{0.53}_{0.47}$ \\ 
		50            &  1.99$^{2.11}_{1.92}$  &   5.65$^{6.15}_{5.36}$ &  3.51$^{3.81}_{3.29}$  &  0.49$^{0.52}_{0.43}$ \\ 
		\hline
		
	\end{tabular}
\end{table}

{ It is important to mention that one major limitation of our model is that it ignores secular changes in the Milky Way potential over time and dynamical friction, which might be important in understanding the evolution of NGC 6362 crossing the inner Galaxy. An in-depth analysis of such dynamical behaviour is beyond the scope of this paper.}

\section{Mass loss rate in NGC 6362}
\label{mass_loss}

The detailed computations of destruction rates of globular clusters in our Galaxy due to the effects of bulge and disk shocking and
dynamical friction, employing the Galactic model \texttt{GravPot16}, will be presented in a future study. However, for the present work we have used destruction rates of the galactic cluster due to dynamical friction and bulge and disk shockings from the literature and added the corresponding destruction rate due to evaporation, to get an estimated value for its total mass loss rate.

\citet{TR} (M+14, hereafter) have computed destruction rates of globular clusters
due to bulge and disk shocking, using a Galactic model which employs a bar component alike the \texttt{GravPot16} model, but with a greater mass, the bar mass ratio being around 1.5. For the orbit of NGC 6362, the kinematic parameters used in the present analysis differ from those used by M+14; however, both models give similar orbits, differing only in the maximum distance $z_{max}$ reached from the Galactic plane, which in our case is around 1.5 times that obtained by M+14. With $t_b$, the characteristic life time due to bulge shocking, M+14 obtain the corresponding present destruction rate $1/t_b = 1.35 \times 10^{-11}$ yr$^{-1}$, using a cluster mass $M_c \sim 10^5$
M$_{\odot}$. With the \texttt{GravPot16} model and the decreased value of $M_c$ in Table 1, $1/t_b$ would be more than the reported value of M+14, but the lower mass of the bar in \texttt{GravPot16} would decrease this value. Thus, we consider the cited value of $1/t_b$ as representative for bulge shocking in our present analysis.

With respect to disk shocking, M+14 obtain the present destruction rate $1/t_d = 2.12 \times 10^{-11}$ yr$^{-1}$, $t_d$ being the corresponding characteristic life time. With the \texttt{GravPot16} model, this value would decrease due to the greater velocity of the cluster when it crosses the Galactic plane as it comes from a greater $z_{max}$ \citep{Spitzer1987}, but with the lower cluster mass given in Table \ref{tab:t1}, $1/t_d$ would increase.

The effect of dynamical friction on globular clusters has been
estimated by \citet{Aguilar1988} taking isotropic velocity dispersion fields in the components of their axisymmetric Galactic models. For NGC 6362, they give $1/t_{df} = 1.4 \times 10^{-12}$ yr$^{-1}$, which is an order of magnitude shorter than $1/t_b$ and $1/t_d$.

To estimate the destruction rate $1/t_{ev}$ due to evaporation, the corresponding life time $t_{ev}$ is computed with $t_{ev} = ft_{rh}$, taking $t_{rh}$ and $f$ given by the equation (7.108) and approximation (7.142) of \citet{Binney2008}. Taking $m$ in that equation as 1 M$_{\odot}$, $M_c = 5.3 \times 10^4$ M$_{\odot}$ (Table 1), and the half-mass radius $r_h = 4.53$ pc (e.g, M+14), the resulting present value for $t_{ev}$, using $f = 40$ is $t_{ev} = 2.4 \times 10^{10}$ yr or an evaporation rate $1/t_{ev} = 4.2 \times 10^{-11}$ yr$^{-1}$.

The sum of $1/t_b$, $1/t_d$, $1/t_{df}$, and $1/t_{ev}$ gives the
total destruction rate $1/t_{tot} = 7.8 \times 10^{-11}$ yr$^{-1}$,
or a present mass loss rate $\dot{M_c} = M_c(1/t_{tot}) = 4.1 \times
10^{-6}$ M$_{\odot}$/yr. To improve this estimate of the mass loss rate, the computation of  $1/t_{df}$ needs to be done with a bar component in the Galactic model, as \texttt{GravPot16} employed here, and taking non-isotropic dispersion fields.

We hypothesise that the mean absolute difference of proper motions in right ascension and declination between the cluster and the 259 possible extended star debris candidates is around 0.5 mas yr$^{-1}$. This gives an approximate mean relative velocity in the plane of the sky of 25 km sec$^{-1}$. With this velocity the stars will move out the vicinity shown in Figure 1 in a time of about 10$^7$ yr. We assume that the star surface density in Figure \ref{fig:f3} is maintained and with the estimated mass loss rate in this interval of time, the cluster loses about 40 M$_{\odot}$. Thus, the majority of the star debris candidates should be low mass stars ($\sim$ 0.15 M$_{\odot}$).

\section{Concluding remarks}
\label{section4}

We have used the {\it Gaia} DR2 information along with the fundamental parameters of the cluster NGC 6362 to search for { possible extended star debris candidates}. We report the identification of 259 potential stellar members of NGC 6362 extending few arc minutes from the edge of the cluster's radius. Both astrometric information and location of these { possible extended star debris} candidates on the CMD are consistent with the cluster membership. Unfortunately, the presently available astrometric information from {\it Gaia} is not sufficient to determine with certainty how many of the stars may be truly { extended star debris} members. Nevertheless, this initial {\it Gaia} DR2 sample significantly contributes to the task of compiling a more thorough census of { possible extended star debris} in the area of the sky around NGC 6362, and portends the promising results to be expected from future spectroscopic follow-up observations.

If the newly discovered objects are part of the main cluster, these results would suggest the presence of an asymmetrically extended stellar material in the outer parts of the cluster whose surface density profile is mainly shaped by evaporation and/or tidal stripping at its current location in the Galaxy; tracing their dynamical evolution in the Milky Way (evaporation and tidal shocking). Also, there is no apparent correlation between the distribution of the newly identified extended star debris candidates and the orbit of the cluster, ruling out any evidence of elongation along the tidal field gradient.\\

The possible extended star debris candidates observed in the cluster can be either due to tidal disruption or dynamical friction or a combined effect of both. Therefore, to find an explanation for these extended star debris candidates, we computed the orbits for the cluster using four different values of $\Omega_{\rm bar}= 35, 40, 45, 50$  km/s/kpc. Half million orbits were computed for different initial conditions considering boxy bar potential perturbations in an inertial reference frame, where the bar is considered at an angle of 20$^{\circ}$ with the line joining Sun and the Galactic center. Earlier, \citet{dana99} also determined the orbital parameters for the cluster, but without the contribution of the bar to the potential. However, the $L_z$ evolution modeled here indicates that the cluster is affected by the bar potential of the Galaxy. Figure~\ref{fig:f3} shows the asymmetric distribution of the { possible extended star debris} candidates along with the orbit of the cluster traced back for 3 Gyr with three different bar speeds. 

Figure~\ref{fig:f9} shows the orbit of the cluster in the meridional Galactic plane and equatorial Galactic plane simulated in the inertial reference frame. It is clear from the figure that the cluster is circulating the inner-disk within a distance of 3 Kpc above and below the disk. As the cluster never enters the bulge of the Galaxy, the dynamical friction experienced by the cluster is negligible, but this cluster has passed through the Galactic disk many times, experiencing a shock every time it crosses the disk. Due to these shocks, many stars must have been stripped away from the cluster. Hence, the observed { extended star debris} candidates can be a result of tidal disruption and shocks from the Galactic disk which happened more than 15.9 Myr. Thanks to the relatively short distance of NGC 6362 and its high release of unbound material during its current disk shocking, we estimate the mass variation to be in the order of $\sim 4.1\times10^{-6}$ M$_{\odot}$ yr$^{-1}$. 

All the raw data used in this work are available through the VizieR Database (I/345/gaia2). Furthermore, in order to facilitate the reproducibility and reuse of our results, we have made available all the data and the source codes available in a public repository\footnote{https://github.com/Fernandez-Trincado/Tidal-debris-Gaia/tree/master/Kundu\%2B2019}.

\begin{figure}
	\begin{center}
		\includegraphics[width=0.55\textwidth,keepaspectratio]{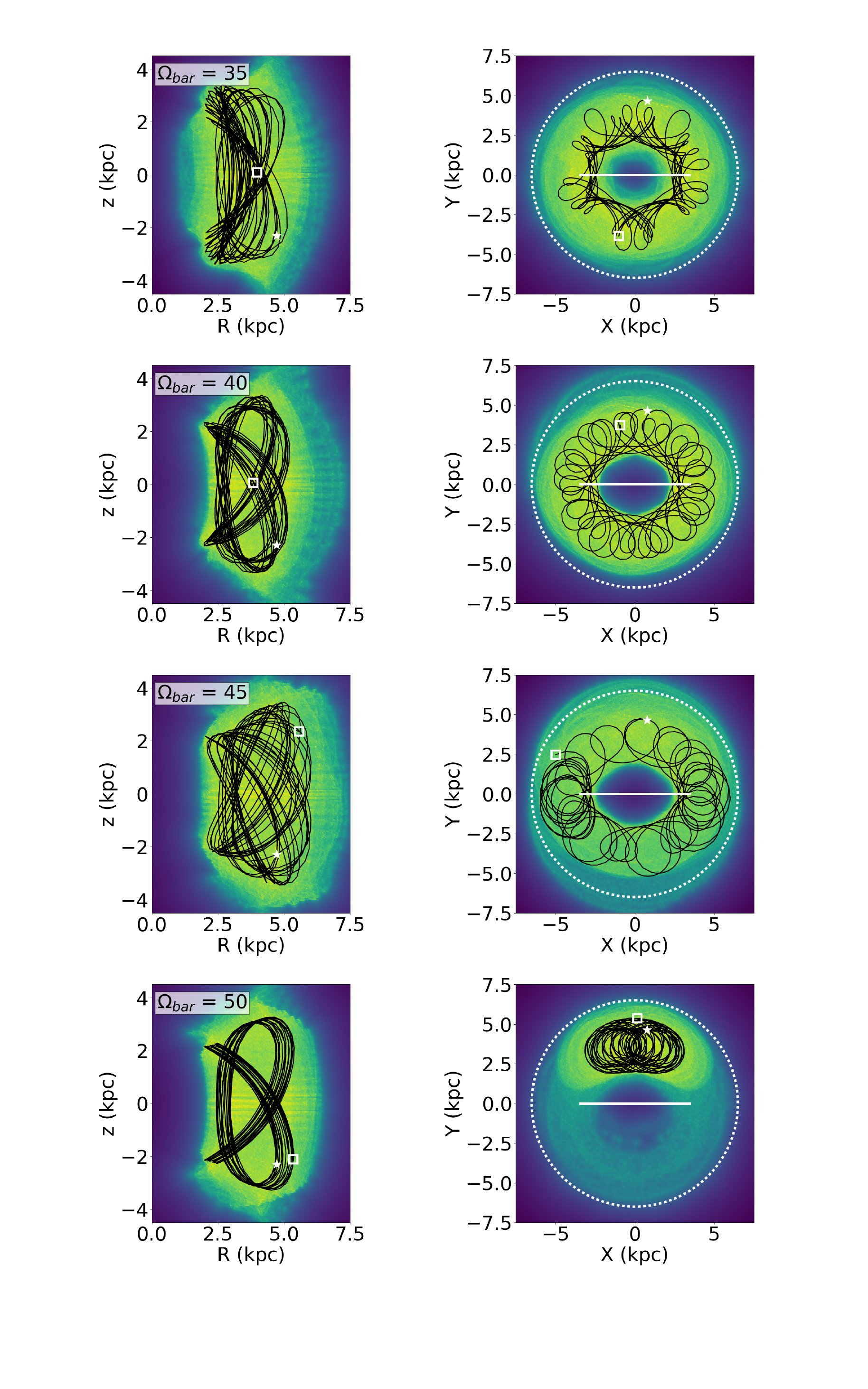}
		\caption{Kernel Density Estimate (KDE) smoothed distribution of simulated orbits employing a Monte Carlo approach, showing the probability densities of the resulting orbits projected on the equatorial (\textit{left}) and meridional (\textit{right}) Galactic planes in the non-inertial reference frame where the bar is at rest. The green and yellow colors correspond to more probable regions of the space, which are crossed more frequently by the simulated orbits. The black line is the orbit of NGC 6362 adopting the central inputs. The small white star marks the present position of the cluster, whereas the white square marks its initial position. In all orbit panels, the white dotted circle show the location of the co-rotation radius (CR), the horizontal white solid line shows the extension of the bar.}
		\label{fig:f9}
	\end{center}
\end{figure}

\begin{figure}
	\begin{center}
		\includegraphics[width=0.55\textwidth,keepaspectratio]{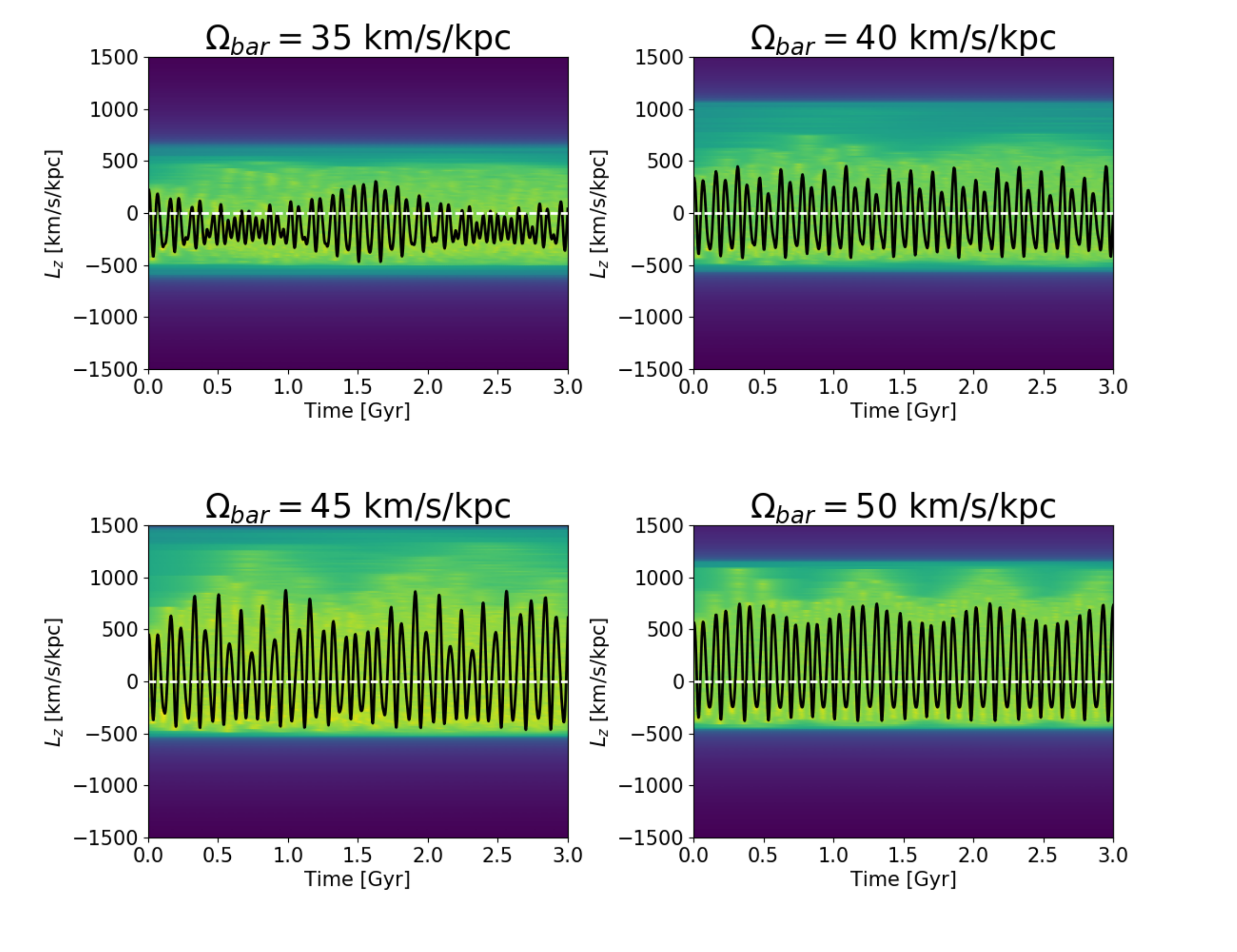}
		\caption{Kernel Density Estimate (KDE) smoothed distribution of the variation of the z-component of the angular momentum ($L_z$) in the inertial frame vs. time for four assumed bar pattern speeds: 35, 40, 45, and 50 km s$^{-1}$ kpc.}
		\label{fig:f8}
	\end{center}
\end{figure}

\section*{Acknowledgements}
 The authors would like to thank the anonymous referee for her/his constructive comments and improvements, making this a better paper. R.K is thankful to the Council of Scientific and Industrial Research, New Delhi, for a Senior Research Fellowship (SRF) (File number: 09/045 (1414)/2016-EMR-I). J.G.F-T is supported by FONDECYT No. 3180210. D.M  gratefully acknowledges support provided by the BASAL  Center for Astrophysics and Associated Technologies (CATA) through grant AFB 170002, and the Ministry for the Economy, Development and Tourism, Programa Iniciativa Cient\'ifica Milenio grant IC120009, awarded to the Millennium Institute of Astrophysics (MAS), and from project Fondecyt No. 1170121.  H.P.S and R.K are thankful to the Council of Scientific and Industrial Research, New Delhi, for the grants-in-aid (Ref. No. 03(1428)/18/EMR-II). R.K and D.M are also very grateful for the hospitality of the Vatican Observatory, where this work was started. E.M acknowledge support from UNAM/PAPIIT grant IN105916.

Funding for the \texttt{GravPot16} software has been provided by the Centre national d'\'etudes spatiales (CNES) through grant 0101973 and UTINAM Institute of the Universit\'e de Franche-Comt\'e, supported by the R\'egion de Franche-Comt\'e and Institut des Sciences de l'Univers (INSU). Simulations have been executed on computers from the Utinam Institute of the Universit\'e de Franche-Comt\'e, supported by the R\'egion de Franche-Comt\'e and Institut des Sciences de l'Univers (INSU), and on the supercomputer facilities of the M\'esocentre de calcul de Franche-Comt\'e. This work has made use of results from the European Space Agency (ESA) space mission {\it Gaia}, the data from which were processed by the {\it Gaia Data Processing and Analysis Consortium} (DPAC). Funding for the DPAC has been provided by national institutions, in particular the institutions participating in the {\it Gaia} Multilateral Agreement. The {\it Gaia} mission website is \url{http: //www.cosmos.esa.int/gaia}. 


\begin{thebibliography}{}
	\makeatletter
	\relax
	\def\mn@urlcharsother{\let\do\@makeother \do\$\do\&\do\#\do\^\do\_\do\%\do\~}
	\def\mn@doi{\begingroup\mn@urlcharsother \@ifnextchar [ {\mn@doi@}
		{\mn@doi@[]}}
	\def\mn@doi@[#1]#2{\def\@tempa{#1}\ifx\@tempa\@empty \href
		{http://dx.doi.org/#2} {doi:#2}\else \href {http://dx.doi.org/#2} {#1}\fi
		\endgroup}
	\def\mn@eprint#1#2{\mn@eprint@#1:#2::\@nil}
	\def\mn@eprint@arXiv#1{\href {http://arxiv.org/abs/#1} {{\tt arXiv:#1}}}
	\def\mn@eprint@dblp#1{\href {http://dblp.uni-trier.de/rec/bibtex/#1.xml}
		{dblp:#1}}
	\def\mn@eprint@#1:#2:#3:#4\@nil{\def\@tempa {#1}\def\@tempb {#2}\def\@tempc
		{#3}\ifx \@tempc \@empty \let \@tempc \@tempb \let \@tempb \@tempa \fi \ifx
		\@tempb \@empty \def\@tempb {arXiv}\fi \@ifundefined
		{mn@eprint@\@tempb}{\@tempb:\@tempc}{\expandafter \expandafter \csname
			mn@eprint@\@tempb\endcsname \expandafter{\@tempc}}}
	
	\bibitem[\protect\citeauthoryear{{Aguilar}, {Hut}  \& {Ostriker}}{{Aguilar}
		et~al.}{1988}]{Aguilar1988}
	{Aguilar} L.,  {Hut} P.,   {Ostriker} J.~P.,  1988, \mn@doi [\apj]
	{10.1086/166961}, \href {http://adsabs.harvard.edu/abs/1988ApJ...335..720A}
	{335, 720}
	
	\bibitem[\protect\citeauthoryear{{Arca-Sedda} \&
		{Capuzzo-Dolcetta}}{{Arca-Sedda} \& {Capuzzo-Dolcetta}}{2014}]{arca14}
	{Arca-Sedda} M.,  {Capuzzo-Dolcetta} R.,  2014, \mn@doi [ApJ]
	{10.1088/0004-637X/785/1/51}, \href
	{http://adsabs.harvard.edu/abs/2014ApJ...785...51A} {785, 51}
	
	\bibitem[\protect\citeauthoryear{{Bailer-Jones}, {Rybizki}, {Fouesneau},
		{Mantelet}  \& {Andrae}}{{Bailer-Jones} et~al.}{2018}]{jones18}
	{Bailer-Jones} C.~A.~L.,  {Rybizki} J.,  {Fouesneau} M.,  {Mantelet} G.,
	{Andrae} R.,  2018, \mn@doi [AJ] {10.3847/1538-3881/aacb21}, \href
	{http://adsabs.harvard.edu/abs/2018AJ....156...58B} {156, 58}
	
	\bibitem[\protect\citeauthoryear{{Balbinot} \& {Gieles}}{{Balbinot} \&
		{Gieles}}{2018}]{balbinot18}
	{Balbinot} E.,  {Gieles} M.,  2018, \mn@doi [MNRAS] {10.1093/mnras/stx2708},
	\href {http://adsabs.harvard.edu/abs/2018MNRAS.474.2479B} {474, 2479}
	
	\bibitem[\protect\citeauthoryear{{Balbinot}, {Santiago}, {da Costa}, {Makler}
		\& {Maia}}{{Balbinot} et~al.}{2011}]{balbinot11}
	{Balbinot} E.,  {Santiago} B.~X.,  {da Costa} L.~N.,  {Makler} M.,   {Maia}
	M.~A.~G.,  2011, \mn@doi [MNRAS] {10.1111/j.1365-2966.2011.19044.x}, \href
	{http://adsabs.harvard.edu/abs/2011MNRAS.416..393B} {416, 393}
	
	\bibitem[\protect\citeauthoryear{{Baumgardt} \& {Hilker}}{{Baumgardt} \&
		{Hilker}}{2018}]{baumgardt18}
	{Baumgardt} H.,  {Hilker} M.,  2018, \mn@doi [MNRAS] {10.1093/mnras/sty1057},
	\href {http://adsabs.harvard.edu/abs/2018MNRAS.478.1520B} {478, 1520}
	
	\bibitem[\protect\citeauthoryear{{Belokurov}, {Evans}, {Irwin}, {Hewett}  \&
		{Wilkinson}}{{Belokurov} et~al.}{2006}]{Belokurov2006}
	{Belokurov} V.,  {Evans} N.~W.,  {Irwin} M.~J.,  {Hewett} P.~C.,   {Wilkinson}
	M.~I.,  2006, \mn@doi [ApJl] {10.1086/500362}, \href
	{http://adsabs.harvard.edu/abs/2006ApJ...637L..29B} {637, L29}
	
	\bibitem[\protect\citeauthoryear{{Binney} \& {Tremaine}}{{Binney} \&
		{Tremaine}}{2008}]{Binney2008}
	{Binney} J.,  {Tremaine} S.,  2008, {Galactic Dynamics: Second Edition}.
	Princeton University Press
	
	\bibitem[\protect\citeauthoryear{{Brunthaler} et~al.,}{{Brunthaler}
		et~al.}{2011}]{Brunthaler2011}
	{Brunthaler} A.,  et~al., 2011, \mn@doi [Astronomische Nachrichten]
	{10.1002/asna.201111560}, \href
	{http://adsabs.harvard.edu/abs/2011AN....332..461B} {332, 461}
	
	\bibitem[\protect\citeauthoryear{{Capuzzo-Dolcetta}}{{Capuzzo-Dolcetta}}{1993}]{capdol93}
	{Capuzzo-Dolcetta} R.,  1993, \mn@doi [ApJ] {10.1086/173189}, \href
	{http://adsabs.harvard.edu/abs/1993ApJ...415..616C} {415, 616}
	
	\bibitem[\protect\citeauthoryear{{Capuzzo-Dolcetta} \&
		{Vicari}}{{Capuzzo-Dolcetta} \& {Vicari}}{2005}]{capdol05b}
	{Capuzzo-Dolcetta} R.,  {Vicari} A.,  2005, \mn@doi [MNRAS]
	{10.1111/j.1365-2966.2004.08433.x}, \href
	{http://adsabs.harvard.edu/abs/2005MNRAS.356..899C} {356, 899}
	
	\bibitem[\protect\citeauthoryear{{Capuzzo Dolcetta}, {Di Matteo}  \&
		{Miocchi}}{{Capuzzo Dolcetta} et~al.}{2005}]{capdel05}
	{Capuzzo Dolcetta} R.,  {Di Matteo} P.,   {Miocchi} P.,  2005, \mn@doi [AJ]
	{10.1086/426006}, \href {http://adsabs.harvard.edu/abs/2005AJ....129.1906C}
	{129, 1906}
	
	\bibitem[\protect\citeauthoryear{{Carretta}, {Bragaglia}, {Gratton},
		{Recio-Blanco}, {Lucatello}, {D'Orazi}  \& {Cassisi}}{{Carretta}
		et~al.}{2010}]{carretta10}
	{Carretta} E.,  {Bragaglia} A.,  {Gratton} R.~G.,  {Recio-Blanco} A.,
	{Lucatello} S.,  {D'Orazi} V.,   {Cassisi} S.,  2010, \mn@doi [AAP]
	{10.1051/0004-6361/200913451}, \href
	{http://adsabs.harvard.edu/abs/2010A%26A...516A..55C} {516, A55}
		
		\bibitem[\protect\citeauthoryear{{Chandrasekhar}}{{Chandrasekhar}}{1943}]{chandrasekhar43}
		{Chandrasekhar} S.,  1943, \mn@doi [ApJ] {10.1086/144517}, \href
		{http://adsabs.harvard.edu/abs/1943ApJ....97..255C} {97, 255}
		
		\bibitem[\protect\citeauthoryear{{Chernoff}, {Kochanek}  \&
			{Shapiro}}{{Chernoff} et~al.}{1986}]{chernoff86}
		{Chernoff} D.~F.,  {Kochanek} C.~S.,   {Shapiro} S.~L.,  1986, \mn@doi [ApJ]
		{10.1086/164591}, \href {http://adsabs.harvard.edu/abs/1986ApJ...309..183C}
		{309, 183}
		
		\bibitem[\protect\citeauthoryear{{Combes}, {Leon}  \& {Meylan}}{{Combes}
			et~al.}{1999}]{combes99}
		{Combes} F.,  {Leon} S.,   {Meylan} G.,  1999, AAP, \href
		{http://adsabs.harvard.edu/abs/1999A%26A...352..149C} {352, 149}
			
			\bibitem[\protect\citeauthoryear{{Dalessandro} et~al.,}{{Dalessandro}
				et~al.}{2014}]{dalessandro14}
			{Dalessandro} E.,  et~al., 2014, \mn@doi [ApJl] {10.1088/2041-8205/791/1/L4},
			\href {http://adsabs.harvard.edu/abs/2014ApJ...791L...4D} {791, L4}
			
			\bibitem[\protect\citeauthoryear{{Dinescu}, {Girard}  \& {van
					Altena}}{{Dinescu} et~al.}{1999}]{dana99}
			{Dinescu} D.~I.,  {Girard} T.~M.,   {van Altena} W.~F.,  1999, \mn@doi [AJ]
			{10.1086/300807}, \href {http://adsabs.harvard.edu/abs/1999AJ....117.1792D}
			{117, 1792}
			
			\bibitem[\protect\citeauthoryear{{Dotter} et~al.,}{{Dotter}
				et~al.}{2010}]{dotter2010}
			{Dotter} A.,  et~al., 2010, \mn@doi [ApJ] {10.1088/0004-637X/708/1/698}, \href
			{http://adsabs.harvard.edu/abs/2010ApJ...708..698D} {708, 698}
			
			\bibitem[\protect\citeauthoryear{{Einasto}}{{Einasto}}{1979}]{Einasto1979}
			{Einasto} J.,  1979, in {Burton} W.~B.,  ed.,  IAU Symposium Vol. 84, The
			Large-Scale Characteristics of the Galaxy. pp 451--458
			
			\bibitem[\protect\citeauthoryear{{Fall} \& {Rees}}{{Fall} \&
				{Rees}}{1977}]{fall77}
			{Fall} S.~M.,  {Rees} M.~J.,  1977, \mn@doi [MNRAS] {10.1093/mnras/181.1.37P},
			\href {http://adsabs.harvard.edu/abs/1977MNRAS.181P..37F} {181, 37P}
			
			\bibitem[\protect\citeauthoryear{{Fall} \& {Rees}}{{Fall} \&
				{Rees}}{1985}]{fall85}
			{Fall} S.~M.,  {Rees} M.~J.,  1985, \mn@doi [ApJ] {10.1086/163585}, \href
			{http://adsabs.harvard.edu/abs/1985ApJ...298...18F} {298, 18}
			
			\bibitem[\protect\citeauthoryear{{Fehlberg}}{{Fehlberg}}{1968}]{fehlberg68}
			{Fehlberg} E.,  1968, NASA Technical Report, p.~315
			
			\bibitem[\protect\citeauthoryear{{Fern{\'a}ndez Trincado}, {Vivas}, {Mateu}  \&
				{Zinn}}{{Fern{\'a}ndez Trincado} et~al.}{2013}]{Fernandez-Trincado2013}
			{Fern{\'a}ndez Trincado} J.~G.,  {Vivas} A.~K.,  {Mateu} C.~E.,   {Zinn} R.,
			2013, \memsai, \href {http://adsabs.harvard.edu/abs/2013MmSAI..84..265F} {84,
				265}
			
			\bibitem[\protect\citeauthoryear{{Fern{\'a}ndez-Trincado}, {Vivas}, {Mateu},
				{Zinn}, {Robin}, {Valenzuela}, {Moreno}  \&
				{Pichardo}}{{Fern{\'a}ndez-Trincado} et~al.}{2015a}]{Fernandez-Trincado2015a}
			{Fern{\'a}ndez-Trincado} J.~G.,  {Vivas} A.~K.,  {Mateu} C.~E.,  {Zinn} R.,
			{Robin} A.~C.,  {Valenzuela} O.,  {Moreno} E.,   {Pichardo} B.,  2015a,
			\mn@doi [\aap] {10.1051/0004-6361/201424899}, \href
			{http://adsabs.harvard.edu/abs/2015A%26A...574A..15F} {574, A15}
				
				\bibitem[\protect\citeauthoryear{{Fern{\'a}ndez-Trincado}
					et~al.,}{{Fern{\'a}ndez-Trincado} et~al.}{2015b}]{Fernandez-Trincado2015b}
				{Fern{\'a}ndez-Trincado} J.~G.,  et~al., 2015b, \mn@doi [\aap]
				{10.1051/0004-6361/201526575}, \href
				{http://adsabs.harvard.edu/abs/2015A%26A...583A..76F} {583, A76}
					
					\bibitem[\protect\citeauthoryear{{Fern{\'a}ndez-Trincado}, {Robin},
						{Reyl{\'e}}, {Vieira}, {Palmer}, {Moreno}, {Valenzuela}  \&
						{Pichardo}}{{Fern{\'a}ndez-Trincado} et~al.}{2016a}]{Fernandez-Trincado2016a}
					{Fern{\'a}ndez-Trincado} J.~G.,  {Robin} A.~C.,  {Reyl{\'e}} C.,  {Vieira} K.,
					{Palmer} M.,  {Moreno} E.,  {Valenzuela} O.,   {Pichardo} B.,  2016a, \mn@doi
					[\mnras] {10.1093/mnras/stw1258}, \href
					{http://adsabs.harvard.edu/abs/2016MNRAS.461.1404F} {461, 1404}
					
					\bibitem[\protect\citeauthoryear{{Fern{\'a}ndez-Trincado}
						et~al.,}{{Fern{\'a}ndez-Trincado} et~al.}{2016b}]{Fernandez-Trincado2016b}
					{Fern{\'a}ndez-Trincado} J.~G.,  et~al., 2016b, \mn@doi [\apj]
					{10.3847/1538-4357/833/2/132}, \href
					{http://adsabs.harvard.edu/abs/2016ApJ...833..132F} {833, 132}
					
					\bibitem[\protect\citeauthoryear{{Fern{\'a}ndez-Trincado}, {Robin}, {Moreno},
						{P{\'e}rez-Villegas}  \& {Pichardo}}{{Fern{\'a}ndez-Trincado}
						et~al.}{2017a}]{Fernandez-Trincado2017b}
					{Fern{\'a}ndez-Trincado} J.~G.,  {Robin} A.~C.,  {Moreno} E.,
					{P{\'e}rez-Villegas} A.,   {Pichardo} B.,  2017a, in {Reyl{\'e}} C.,  {Di
						Matteo} P.,  {Herpin} F.,  {Lagadec} E.,  {Lan{\c c}on} A.,  {Meliani} Z.,
					{Royer} F.,  eds, SF2A-2017: Proceedings of the Annual meeting of the French
					Society of Astronomy and Astrophysics. pp 193--198 (\mn@eprint {arXiv}
					{1708.05742})
					
					\bibitem[\protect\citeauthoryear{{Fern{\'a}ndez-Trincado}, {Geisler}, {Moreno},
						{Zamora}, {Robin}  \& {Villanova}}{{Fern{\'a}ndez-Trincado}
						et~al.}{2017b}]{Fernandez-Trincado2017c}
					{Fern{\'a}ndez-Trincado} J.~G.,  {Geisler} D.,  {Moreno} E.,  {Zamora} O.,
					{Robin} A.~C.,   {Villanova} S.,  2017b, in {Reyl{\'e}} C.,  {Di Matteo} P.,
					{Herpin} F.,  {Lagadec} E.,  {Lan{\c c}on} A.,  {Meliani} Z.,   {Royer} F.,
					eds, SF2A-2017: Proceedings of the Annual meeting of the French Society of
					Astronomy and Astrophysics. pp 199--202 (\mn@eprint {arXiv} {1710.07433})
					
					\bibitem[\protect\citeauthoryear{{Fern{\'a}ndez-Trincado}
						et~al.,}{{Fern{\'a}ndez-Trincado} et~al.}{2017c}]{Fernandez-Trincado2017a}
					{Fern{\'a}ndez-Trincado} J.~G.,  et~al., 2017c, \mn@doi [\apjl]
					{10.3847/2041-8213/aa8032}, \href
					{http://adsabs.harvard.edu/abs/2017ApJ...846L...2F} {846, L2}
					
					\bibitem[\protect\citeauthoryear{{Fern{\'a}ndez-Trincado}
						et~al.,}{{Fern{\'a}ndez-Trincado} et~al.}{2019a}]{Fernandez-Trincado2019a}
					{Fern{\'a}ndez-Trincado} J.~G.,  et~al., 2019a, arXiv e-prints, \href
					{http://adsabs.harvard.edu/abs/2019arXiv190210635F} {}
					
					\bibitem[\protect\citeauthoryear{{Fern{\'a}ndez-Trincado}, {Ortigoza-Urdaneta},
						{Moreno}, {P{\'e}rez-Villegas}  \& {Soto}}{{Fern{\'a}ndez-Trincado}
						et~al.}{2019c}]{Fernandez-Trincado2019c}
					{Fern{\'a}ndez-Trincado} J.~G.,  {Ortigoza-Urdaneta} M.,  {Moreno} E.,
					{P{\'e}rez-Villegas} A.,   {Soto} M.,  2019c, arXiv e-prints, \href
					{http://adsabs.harvard.edu/abs/2019arXiv190405370F} {}
					
					\bibitem[\protect\citeauthoryear{{Fern{\'a}ndez-Trincado}
						et~al.,}{{Fern{\'a}ndez-Trincado} et~al.}{2019b}]{Fernandez-Trincado2019d}
					{Fern{\'a}ndez-Trincado} J.~G.,  et~al., 2019b, arXiv e-prints, \href
					{http://adsabs.harvard.edu/abs/2019arXiv190405884F} {}
					
					\bibitem[\protect\citeauthoryear{{Fern{\'a}ndez-Trincado}, {Beers}, {Tang},
						{Moreno}, {P{\'e}rez-Villegas}  \&
						{Ortigoza-Urdaneta}}{{Fern{\'a}ndez-Trincado}
						et~al.}{2019d}]{Fernandez-Trincado2019b}
					{Fern{\'a}ndez-Trincado} J.~G.,  {Beers} T.~C.,  {Tang} B.,  {Moreno} E.,
					{P{\'e}rez-Villegas} A.,   {Ortigoza-Urdaneta} M.,  2019d, \mn@doi [\mnras]
					{10.1093/mnras/stz1848}, \href
					{https://ui.adsabs.harvard.edu/abs/2019MNRAS.488.2864F} {488, 2864}
					
					\bibitem[\protect\citeauthoryear{{Fern{\'a}ndez-Trincado}
						et~al.,}{{Fern{\'a}ndez-Trincado} et~al.}{2019e}]{Fernandez-Trincado2018}
					{Fern{\'a}ndez-Trincado} J.~G.,  et~al., 2019e, \mn@doi [\aap]
					{10.1051/0004-6361/201834391}, \href
					{https://ui.adsabs.harvard.edu/abs/2019A%26A...627A.178F} {627, A178}
						
						\bibitem[\protect\citeauthoryear{{Gaia Collaboration} et~al.,}{{Gaia
								Collaboration} et~al.}{2018}]{gaiadr2}
						{Gaia Collaboration} et~al., 2018, \mn@doi [AAP] {10.1051/0004-6361/201833051},
						\href {http://adsabs.harvard.edu/abs/2018A%26A...616A...1G} {616, A1}
							
							\bibitem[\protect\citeauthoryear{{Gnedin} \& {Ostriker}}{{Gnedin} \&
								{Ostriker}}{1997}]{gnedin97}
							{Gnedin} O.~Y.,  {Ostriker} J.~P.,  1997, \mn@doi [ApJ] {10.1086/303441}, \href
							{http://adsabs.harvard.edu/abs/1997ApJ...474..223G} {474, 223}
							
							\bibitem[\protect\citeauthoryear{{Grillmair} \& {Johnson}}{{Grillmair} \&
								{Johnson}}{2006}]{Grillmair2006}
							{Grillmair} C.~J.,  {Johnson} R.,  2006, \mn@doi [ApJl] {10.1086/501439}, \href
							{http://adsabs.harvard.edu/abs/2006ApJ...639L..17G} {639, L17}
							
							\bibitem[\protect\citeauthoryear{{Grillmair} \& {Mattingly}}{{Grillmair} \&
								{Mattingly}}{2010}]{Grillmair10}
							{Grillmair} C.~J.,  {Mattingly} S.,  2010, in American Astronomical Society
							Meeting Abstracts \#216. p.~833
							
							\bibitem[\protect\citeauthoryear{{Hozumi} \& {Burkert}}{{Hozumi} \&
								{Burkert}}{2015}]{hozumi15}
							{Hozumi} S.,  {Burkert} A.,  2015, \mn@doi [MNRAS] {10.1093/mnras/stu2318},
							\href {http://adsabs.harvard.edu/abs/2015MNRAS.446.3100H} {446, 3100}
							
							\bibitem[\protect\citeauthoryear{{Jordi} \& {Grebel}}{{Jordi} \&
								{Grebel}}{2010}]{jordi2010}
							{Jordi} K.,  {Grebel} E.~K.,  2010, \mn@doi [AAP]
							{10.1051/0004-6361/201014392}, \href
							{http://adsabs.harvard.edu/abs/2010A%26A...522A..71J} {522, A71}
								
								\bibitem[\protect\citeauthoryear{{King}}{{King}}{1962}]{king62}
								{King} I.,  1962, \mn@doi [AJ] {10.1086/108756}, \href
								{http://adsabs.harvard.edu/abs/1962AJ.....67..471K} {67, 471}
								
								\bibitem[\protect\citeauthoryear{{Knierman}, {Scowen}, {Veach}, {Groppi},
									{Mullan}, {Konstantopoulos}, {Knezek}  \& {Charlton}}{{Knierman}
									et~al.}{2013}]{knierman13}
								{Knierman} K.~A.,  {Scowen} P.,  {Veach} T.,  {Groppi} C.,  {Mullan} B.,
								{Konstantopoulos} I.,  {Knezek} P.~M.,   {Charlton} J.,  2013, \mn@doi [ApJ]
								{10.1088/0004-637X/774/2/125}, \href
								{http://adsabs.harvard.edu/abs/2013ApJ...774..125K} {774, 125}
								
								\bibitem[\protect\citeauthoryear{{Kunder} et~al.,}{{Kunder}
									et~al.}{2014}]{kunder2014}
								{Kunder} A.,  et~al., 2014, \mn@doi [AAP] {10.1051/0004-6361/201424113}, \href
								{http://adsabs.harvard.edu/abs/2014A%26A...572A..30K} {572, A30}
									
									\bibitem[\protect\citeauthoryear{{Kunder} et~al.,}{{Kunder}
										et~al.}{2018}]{Kunder18}
									{Kunder} A.,  et~al., 2018, \mn@doi [AJ] {10.3847/1538-3881/aab42d}, \href
									{http://adsabs.harvard.edu/abs/2018AJ....155..171K} {155, 171}
									
									\bibitem[\protect\citeauthoryear{{Kundu}, {Minniti}  \& {Singh}}{{Kundu}
										et~al.}{2019}]{Kundu2019}
									{Kundu} R.,  {Minniti} D.,   {Singh} H.~P.,  2019, \mn@doi [\mnras]
									{10.1093/mnras/sty3239}, \href
									{http://adsabs.harvard.edu/abs/2019MNRAS.483.1737K} {483, 1737}
									
									\bibitem[\protect\citeauthoryear{{K{\"u}pper}, {Lane}  \&
										{Heggie}}{{K{\"u}pper} et~al.}{2012}]{kupper12}
									{K{\"u}pper} A.~H.~W.,  {Lane} R.~R.,   {Heggie} D.~C.,  2012, \mn@doi [MNRAS]
									{10.1111/j.1365-2966.2011.20242.x}, \href
									{http://adsabs.harvard.edu/abs/2012MNRAS.420.2700K} {420, 2700}
									
									\bibitem[\protect\citeauthoryear{{Kuzma}, {Da Costa}, {Keller}  \&
										{Maunder}}{{Kuzma} et~al.}{2015}]{kuzma15}
									{Kuzma} P.~B.,  {Da Costa} G.~S.,  {Keller} S.~C.,   {Maunder} E.,  2015,
									\mn@doi [MNRAS] {10.1093/mnras/stu2343}, \href
									{http://adsabs.harvard.edu/abs/2015MNRAS.446.3297K} {446, 3297}
									
									\bibitem[\protect\citeauthoryear{{Lagarde}, {Robin}, {Reyl{\'e}}  \&
										{Nasello}}{{Lagarde} et~al.}{2017}]{Lagarde2017}
									{Lagarde} N.,  {Robin} A.~C.,  {Reyl{\'e}} C.,   {Nasello} G.,  2017, \mn@doi
									[\aap] {10.1051/0004-6361/201630253}, \href
									{http://adsabs.harvard.edu/abs/2017A%26A...601A..27L} {601, A27}
										
										\bibitem[\protect\citeauthoryear{{Leon}, {Meylan}  \& {Combes}}{{Leon}
											et~al.}{2000}]{Leon2000}
										{Leon} S.,  {Meylan} G.,   {Combes} F.,  2000, AAP, \href
										{http://adsabs.harvard.edu/abs/2000A%26A...359..907L} {359, 907}
											
											\bibitem[\protect\citeauthoryear{{Lotz}, {Telford}, {Ferguson}, {Miller},
												{Stiavelli}  \& {Mack}}{{Lotz} et~al.}{2001}]{lotz01}
											{Lotz} J.~M.,  {Telford} R.,  {Ferguson} H.~C.,  {Miller} B.~W.,  {Stiavelli}
											M.,   {Mack} J.,  2001, \mn@doi [ApJ] {10.1086/320545}, \href
											{http://adsabs.harvard.edu/abs/2001ApJ...552..572L} {552, 572}
											
											\bibitem[\protect\citeauthoryear{{Mackereth} et~al.,}{{Mackereth}
												et~al.}{2019}]{Mackereth2019}
											{Mackereth} J.~T.,  et~al., 2019, \mn@doi [\mnras] {10.1093/mnras/sty2955},
											\href {https://ui.adsabs.harvard.edu/abs/2019MNRAS.482.3426M} {482, 3426}
											
											\bibitem[\protect\citeauthoryear{{Majewski} et~al.,}{{Majewski}
												et~al.}{2012a}]{majewski12}
											{Majewski} S.~R.,  et~al., 2012a, in American Astronomical Society Meeting
											Abstracts \#219. p. 410.05
											
											\bibitem[\protect\citeauthoryear{{Majewski}, {Nidever}, {Smith}, {Damke},
												{Kunkel}, {Patterson}, {Bizyaev}  \& {Garc{\'{\i}}a P{\'e}rez}}{{Majewski}
												et~al.}{2012b}]{Majewski12a}
											{Majewski} S.~R.,  {Nidever} D.~L.,  {Smith} V.~V.,  {Damke} G.~J.,  {Kunkel}
											W.~E.,  {Patterson} R.~J.,  {Bizyaev} D.,   {Garc{\'{\i}}a P{\'e}rez} A.~E.,
											2012b, \mn@doi [ApJl] {10.1088/2041-8205/747/2/L37}, \href
											{http://adsabs.harvard.edu/abs/2012ApJ...747L..37M} {747, L37}
											
											\bibitem[\protect\citeauthoryear{{Majewski}, {APOGEE Team}  \& {APOGEE-2
													Team}}{{Majewski} et~al.}{2016}]{Majewski2016}
											{Majewski} S.~R.,  {APOGEE Team}  {APOGEE-2 Team} 2016, \mn@doi [Astronomische
											Nachrichten] {10.1002/asna.201612387}, \href
											{https://ui.adsabs.harvard.edu/abs/2016AN....337..863M} {337, 863}
											
											\bibitem[\protect\citeauthoryear{{Marchetti}, {Rossi}  \& {Brown}}{{Marchetti}
												et~al.}{2018}]{Marchetti18}
											{Marchetti} T.,  {Rossi} E.~M.,   {Brown} A.~G.~A.,  2018, \mn@doi [MNRAS]
											{10.1093/mnras/sty2592}, \href
											{http://adsabs.harvard.edu/abs/2018MNRAS.tmp.2466M} {}
											
											\bibitem[\protect\citeauthoryear{{Massari} et~al.,}{{Massari}
												et~al.}{2017}]{massari17}
											{Massari} D.,  et~al., 2017, \mn@doi [MNRAS] {10.1093/mnras/stx549}, \href
											{http://adsabs.harvard.edu/abs/2017MNRAS.468.1249M} {468, 1249}
											
											\bibitem[\protect\citeauthoryear{{Meylan} \& {Heggie}}{{Meylan} \&
												{Heggie}}{1997}]{meylan97}
											{Meylan} G.,  {Heggie} D.~C.,  1997, \mn@doi [AAPr] {10.1007/s001590050008},
											\href {http://adsabs.harvard.edu/abs/1997A%26ARv...8....1M} {8, 1}
												
												\bibitem[\protect\citeauthoryear{{Minniti}, {Fern{\'a}ndez-Trincado}, {Ripepi},
													{Alonso-Garc{\'{\i}}a}, {Contreras Ramos}  \& {Marconi}}{{Minniti}
													et~al.}{2018}]{Minniti18}
												{Minniti} D.,  {Fern{\'a}ndez-Trincado} J.~G.,  {Ripepi} V.,
												{Alonso-Garc{\'{\i}}a} J.,  {Contreras Ramos} R.,   {Marconi} M.,  2018,
												\mn@doi [\apjl] {10.3847/2041-8213/aaf1cd}, \href
												{http://adsabs.harvard.edu/abs/2018ApJ...869L..10M} {869, L10}
												
												\bibitem[\protect\citeauthoryear{{Moreno}, {Pichardo}  \&
													{Vel{\'a}zquez}}{{Moreno} et~al.}{2014}]{TR}
												{Moreno} E.,  {Pichardo} B.,   {Vel{\'a}zquez} H.,  2014, \mn@doi [ApJ]
												{10.1088/0004-637X/793/2/110}, \href
												{http://adsabs.harvard.edu/abs/2014ApJ...793..110M} {793, 110}
												
												\bibitem[\protect\citeauthoryear{{Mulder}}{{Mulder}}{1983}]{mulder83}
												{Mulder} W.~A.,  1983, AAP, \href
												{http://adsabs.harvard.edu/abs/1983A%26A...117....9M} {117, 9}
													
													\bibitem[\protect\citeauthoryear{{Mulia} \& {Chandar}}{{Mulia} \&
														{Chandar}}{2014}]{mulia14}
													{Mulia} A.,  {Chandar} R.,  2014, in American Astronomical Society Meeting
													Abstracts \#223. p. 442.34
													
													\bibitem[\protect\citeauthoryear{{Murali} \& {Weinberg}}{{Murali} \&
														{Weinberg}}{1997}]{murali1997}
													{Murali} C.,  {Weinberg} M.~D.,  1997, \mn@doi [MNRAS]
													{10.1093/mnras/291.4.717}, \href
													{http://adsabs.harvard.edu/abs/1997MNRAS.291..717M} {291, 717}
													
													\bibitem[\protect\citeauthoryear{{Myeong}, {Jerjen}, {Mackey}  \& {Da
															Costa}}{{Myeong} et~al.}{2017}]{myeong17}
													{Myeong} G.~C.,  {Jerjen} H.,  {Mackey} D.,   {Da Costa} G.~S.,  2017, \mn@doi
													[ApJl] {10.3847/2041-8213/aa6fb4}, \href
													{http://adsabs.harvard.edu/abs/2017ApJ...840L..25M} {840, L25}
													
													\bibitem[\protect\citeauthoryear{{Myeong}, {Evans}, {Belokurov}, {Sanders}  \&
														{Koposov}}{{Myeong} et~al.}{2018}]{Myeong18}
													{Myeong} G.~C.,  {Evans} N.~W.,  {Belokurov} V.,  {Sanders} J.~L.,   {Koposov}
													S.~E.,  2018, \mn@doi [MNRAS] {10.1093/mnras/sty1403}, \href
													{http://adsabs.harvard.edu/abs/2018MNRAS.478.5449M} {478, 5449}
													
													\bibitem[\protect\citeauthoryear{{Navarrete}, {Belokurov}  \&
														{Koposov}}{{Navarrete} et~al.}{2017}]{navarrete17}
													{Navarrete} C.,  {Belokurov} V.,   {Koposov} S.~E.,  2017, \mn@doi [ApJl]
													{10.3847/2041-8213/aa72e1}, \href
													{http://adsabs.harvard.edu/abs/2017ApJ...841L..23N} {841, L23}
													
													\bibitem[\protect\citeauthoryear{{Niederste-Ostholt}, {Belokurov}, {Evans},
														{Koposov}, {Gieles}  \& {Irwin}}{{Niederste-Ostholt}
														et~al.}{2010}]{ostholt10}
													{Niederste-Ostholt} M.,  {Belokurov} V.,  {Evans} N.~W.,  {Koposov} S.,
													{Gieles} M.,   {Irwin} M.~J.,  2010, \mn@doi [MNRAS]
													{10.1111/j.1745-3933.2010.00931.x}, \href
													{http://adsabs.harvard.edu/abs/2010MNRAS.408L..66N} {408, L66}
													
													\bibitem[\protect\citeauthoryear{{Odenkirchen} et~al.,}{{Odenkirchen}
														et~al.}{2001}]{Odenkirchen2001}
													{Odenkirchen} M.,  et~al., 2001, \mn@doi [ApJl] {10.1086/319095}, \href
													{http://adsabs.harvard.edu/abs/2001ApJ...548L.165O} {548, L165}
													
													\bibitem[\protect\citeauthoryear{{Robin}, {Reyl{\'e}}, {Derri{\`e}re}  \&
														{Picaud}}{{Robin} et~al.}{2003}]{Robin2003}
													{Robin} A.~C.,  {Reyl{\'e}} C.,  {Derri{\`e}re} S.,   {Picaud} S.,  2003,
													\mn@doi [\aap] {10.1051/0004-6361:20031117}, \href
													{http://adsabs.harvard.edu/abs/2003A%26A...409..523R} {409, 523}
														
														\bibitem[\protect\citeauthoryear{{Robin}, {Marshall}, {Schultheis}  \&
															{Reyl{\'e}}}{{Robin} et~al.}{2012}]{Robin2012}
														{Robin} A.~C.,  {Marshall} D.~J.,  {Schultheis} M.,   {Reyl{\'e}} C.,  2012,
														\mn@doi [AAP] {10.1051/0004-6361/201116512}, \href
														{http://adsabs.harvard.edu/abs/2012A%26A...538A.106R} {538, A106}
															
															\bibitem[\protect\citeauthoryear{{Robin}, {Reyl{\'e}}, {Fliri}, {Czekaj},
																{Robert}  \& {Martins}}{{Robin} et~al.}{2014}]{Robin2014}
															{Robin} A.~C.,  {Reyl{\'e}} C.,  {Fliri} J.,  {Czekaj} M.,  {Robert} C.~P.,
															{Martins} A.~M.~M.,  2014, \mn@doi [\aap] {10.1051/0004-6361/201423415},
															\href {http://adsabs.harvard.edu/abs/2014A%26A...569A..13R} {569, A13}
																
																\bibitem[\protect\citeauthoryear{{Robin}, {Bienaym{\'e}},
																	{Fern{\'a}ndez-Trincado}  \& {Reyl{\'e}}}{{Robin} et~al.}{2017}]{Robin2017}
																{Robin} A.~C.,  {Bienaym{\'e}} O.,  {Fern{\'a}ndez-Trincado} J.~G.,
																{Reyl{\'e}} C.,  2017, \mn@doi [\aap] {10.1051/0004-6361/201630217}, \href
																{https://ui.adsabs.harvard.edu/abs/2017A%26A...605A...1R} {605, A1}
																	
																	\bibitem[\protect\citeauthoryear{{Rodruck} et~al.,}{{Rodruck}
																		et~al.}{2016}]{rodruck16}
																	{Rodruck} M.,  et~al., 2016, \mn@doi [MNRAS] {10.1093/mnras/stw1294}, \href
																	{http://adsabs.harvard.edu/abs/2016MNRAS.461...36R} {461, 36}
																	
																	\bibitem[\protect\citeauthoryear{{Sollima}, {Mart{\'{\i}}nez-Delgado},
																		{Valls-Gabaud}  \& {Pe{\~n}arrubia}}{{Sollima} et~al.}{2011}]{sollima11}
																	{Sollima} A.,  {Mart{\'{\i}}nez-Delgado} D.,  {Valls-Gabaud} D.,
																	{Pe{\~n}arrubia} J.,  2011, \mn@doi [ApJ] {10.1088/0004-637X/726/1/47}, \href
																	{http://adsabs.harvard.edu/abs/2011ApJ...726...47S} {726, 47}
																	
																	\bibitem[\protect\citeauthoryear{{Spitzer}}{{Spitzer}}{1987}]{Spitzer1987}
																	{Spitzer} L.,  1987, {Dynamical evolution of globular clusters}
																	
																	\bibitem[\protect\citeauthoryear{{Torres-Flores}, {de Oliveira}, {de Mello},
																		{Scarano}  \& {Urrutia-Viscarra}}{{Torres-Flores} et~al.}{2012}]{flores12}
																	{Torres-Flores} S.,  {de Oliveira} C.~M.,  {de Mello} D.~F.,  {Scarano} S.,
																	{Urrutia-Viscarra} F.,  2012, \mn@doi [MNRAS]
																	{10.1111/j.1365-2966.2012.20589.x}, \href
																	{http://adsabs.harvard.edu/abs/2012MNRAS.421.3612T} {421, 3612}
																	
																	\bibitem[\protect\citeauthoryear{{Tremaine} \& {Weinberg}}{{Tremaine} \&
																		{Weinberg}}{1984}]{tremaine84}
																	{Tremaine} S.,  {Weinberg} M.~D.,  1984, \mn@doi [MNRAS]
																	{10.1093/mnras/209.4.729}, \href
																	{http://adsabs.harvard.edu/abs/1984MNRAS.209..729T} {209, 729}
																	
																	\bibitem[\protect\citeauthoryear{{Tremaine}, {Ostriker}  \&
																		{Spitzer}}{{Tremaine} et~al.}{1975}]{tremaine75}
																	{Tremaine} S.~D.,  {Ostriker} J.~P.,   {Spitzer} Jr. L.,  1975, \mn@doi [ApJ]
																	{10.1086/153422}, \href {http://adsabs.harvard.edu/abs/1975ApJ...196..407T}
																	{196, 407}
																	
																	\bibitem[\protect\citeauthoryear{{Vasiliev}}{{Vasiliev}}{2019}]{Vasiliev18}
																	{Vasiliev} E.,  2019, \mn@doi [\mnras] {10.1093/mnras/stz171}, \href
																	{http://adsabs.harvard.edu/abs/2019MNRAS.484.2832V} {484, 2832}
																	
																	\bibitem[\protect\citeauthoryear{{Vesperini} \& {Heggie}}{{Vesperini} \&
																		{Heggie}}{1997}]{vesperini97}
																	{Vesperini} E.,  {Heggie} D.~C.,  1997, \mn@doi [MNRAS]
																	{10.1093/mnras/289.4.898}, \href
																	{http://adsabs.harvard.edu/abs/1997MNRAS.289..898V} {289, 898}
																	
																	\bibitem[\protect\citeauthoryear{{Weinberg}}{{Weinberg}}{1994}]{weinberg94}
																	{Weinberg} M.~D.,  1994, \mn@doi [AJ] {10.1086/117163}, \href
																	{http://adsabs.harvard.edu/abs/1994AJ....108.1414W} {108, 1414}
																	
																	\bibitem[\protect\citeauthoryear{{White}}{{White}}{1983}]{white83}
																	{White} S.~D.~M.,  1983, \mn@doi [ApJ] {10.1086/161425}, \href
																	{http://adsabs.harvard.edu/abs/1983ApJ...274...53W} {274, 53}
																	
																	\bibitem[\protect\citeauthoryear{{Zasowski} et~al.,}{{Zasowski}
																		et~al.}{2017}]{Zasowski2017}
																	{Zasowski} G.,  et~al., 2017, \mn@doi [\aj] {10.3847/1538-3881/aa8df9}, \href
																	{https://ui.adsabs.harvard.edu/abs/2017AJ....154..198Z} {154, 198}
																	
																	\makeatother
																\end{thebibliography}



\begin{table*}
	\centering
		\setlength{\tabcolsep}{3.5mm}  
	\caption{{ Possible extended star debris} candidates of NGC 6362 from {\it Gaia} DR2.}
	\label{tab:tdata}
	\begin{tabular}{cccccccc} 
		\hline
ID	&	$\alpha$	&	$\delta$	&	$\mu_{\alpha}$ &  $\mu_{\delta}$  &  $G$	&	$G_{BP}$	&	$G_{RP}$\\
&	($^{\circ}$)	&($^{\circ}$)	&	(mas/yr) &  (mas/yr)  &  (mag)&	(mag)	&	(mag)\\
\hline
5810760588765404672  &  263.950  &  -68.123  &   -4.546  $\pm$    0.150  &   -4.631  $\pm$    0.202  &   17.932  &   18.331  &   17.362  \\
5810766331142949376  &  264.066  &  -68.122  &   -4.999  $\pm$    0.129  &   -5.362  $\pm$    0.166  &   17.641  &   18.063  &   17.058  \\
5810767636813138304  &  264.222  &  -68.057  &   -5.224  $\pm$    0.028  &   -4.442  $\pm$    0.037  &   14.585  &   15.153  &   13.881  \\
5811490290829953280  &  263.323  &  -68.172  &   -6.203  $\pm$    0.163  &   -4.243  $\pm$    0.230  &   18.149  &   18.547  &   17.623  \\
5811498086186458752  &  262.698  &  -68.193  &   -4.615  $\pm$    0.167  &   -4.520  $\pm$    0.222  &   18.143  &   18.515  &   17.598  \\
5811500457010583552  &  263.041  &  -68.198  &   -5.140  $\pm$    0.218  &   -5.427  $\pm$    0.318  &   18.662  &   18.993  &   18.140  \\
5811501212924838144  &  262.897  &  -68.194  &   -5.095  $\pm$    0.158  &   -5.473  $\pm$    0.233  &   18.115  &   18.477  &   17.589  \\
5811501973141042304  &  263.066  &  -68.188  &   -4.907  $\pm$    0.087  &   -5.626  $\pm$    0.136  &   17.035  &   17.492  &   16.410  \\
\hline
\end{tabular}\\
	\raggedright {{ Note:} Table 3 is published in its entirety in a public repository at \url{https://github.com/Fernandez-Trincado/Tidal-debris-Gaia/tree/master/Kundu\%2B2019}. A portion is shown here for guidance regarding its form and content.
 }
\end{table*}


\end{document}